# Building the Principle of Thermoelectric *ZT* Enhancement

Shuang Tang [1]* and Mildred S. Dresselhaus [2,3]

[1] Department of Materials Science and Engineering, Massachusetts Institute of Technology, Cambridge, MA, 02139, USA ( tangs@mit.edu)

[2] Department of Electrical Engineering and Computer Science, and [3] Department of Physics, Massachusetts Institute of Technology, Cambridge MA 02139, USA

**Thermoelectrics is the topic on convertion between heat flow and electricity, the performance of which is characterized by the dimensionless figure of merit $ZT=\sigma S^2 T/\kappa$, where $\sigma$, $S$, $T$ and $\kappa$ are the electrical conductivity, Seebeck coefficient, temperature and thermal conductivity, respectively. $ZT$ was believed to have an upper limit of 1 [1] until the recent two decades, many novel approaches have pushed the upper limit sequentially, including the proposals of low-dimensionalization [2], sharp density of states [3], superlattice [4-6], resonant states [7], nanocomposites [8,9], pipe-shaped Fermi surface [10], etc. However, the problem of enhancing $ZT$ to an industrially competitive level is not solved yet. The strong correlation between $\sigma$, $S$ and $\kappa$ results in significant difficulties on the optimization problem. Materials with large $\sigma$ always tend to have large $\kappa$ and small $S$, which kills $ZT$, and vice versa. Before this present paper, there has not been a framework of standard principles on $ZT$ enhancing, including how to choose thermoelectric materials among a pool of candidates and how to improve the thermoelectric behavior within a single material system. Our goal is to build a systematic framework of principles for thermoelectric figure of merit $ZT$ enhancing, regarding choosing and improving materials. We have pointed out that the current expression of $ZT$ is actually the obstacle of the $ZT$ enhancement problem. We have then find the variables that really decides and really enhance $ZT$ by proposing the idea of splitting $ZT$ into two pseudo-$ZT$s, after which we have succeeded in building the principles of ZT enhancing, regarding dimension of ma-**

**terials, dispersion relation forms, carrier scattering mechanism, density of states, band asymmetry and band gap.**

This present paper aims on exploring why this enhancing *ZT* problem is barely soluble in its current form, and how it should be re-formed to be soluble. We start from pointing out the mathematical fact that it is the way of expressing *ZT* by the variables of $\sigma$, *S*, and $\kappa$ that makes the optimization task scarcely achievable. To discuss the way of expressing *ZT* that can make the optimization problem mathematically soluble, we introduce the idea of splitting *ZT* into two pseudo-*ZT*s: one characterizes the influence of electronic behaviors to *ZT*; the other scales the influence of the lattice thermal conductivity to *ZT*. After introducing this scheme of pseudo-*ZT*s, we can then make it clear what variables really decides *ZT*. Then we explore the most important question on what variables really enhance *ZT*.

Various proposals have been made to improve *ZT*, by choosing or changing dimensions of materials, dispersion relation forms of band valleys, number of band valleys, carrier scattering mechanism, density of states, band asymmetry, band gap, etc, within one materials system or among a pool of materials candidates. It has been puzzling that these various proposals do not always work on improving *ZT*. We will answer this puzzle below while discussing what are the variables that really enhance *ZT*, and then give the physical guidance on how to choose the correct proposals under different conditions.

Firstly, we point out that when *ZT* is written as $ZT=\sigma S^2 T/\kappa$, we are already making the problem barely soluble. Enhancing *ZT* is basically an optimization problem in mathematics. However, it is not allowed to change one of the three variables ($\sigma$, *S* and $\kappa$) while having the other two fixed, e.g. it is unable to enhance *ZT* by increasing $\sigma$ and keeping *S* and $\kappa$ unchanged, because $\sigma$, *S* and $\kappa$ are highly correlated. Thus, our first goal is to find a way of expressing *ZT* by variables that are independent or weakly correlated to each other.

To achieve this goal, we introduce the idea of splitting *ZT* into two pseudo-*ZT*s: $zt_e$ and $zt_L$. Three generally accepted assumptions are made before the derivation of pseudo-*ZT*s: (1) the transport can be described by the Boltzmann equations [11,12], i.e.

$$\sigma = e^2 \cdot I_0 \tag{1}$$

$$S = \frac{k_B}{e} \frac{I_1}{I_0} \tag{2}$$

$$\kappa_e = T \cdot k_B^2 I_2 - T\sigma S^2 \tag{3}$$

where $I_r = \int (-\partial f_0/\partial E) \Xi(E)((E-E_f)/k_B \cdot T)^r dE$ and $\Xi(E)$ is the transport distribution function. (2) The transport distribution function $\Xi(E)$ can be characterized by the "exponent law" [11,12], i.e.

$$\Xi(E) = g_e \cdot a_n \cdot \left(\frac{E}{k_B T}\right)^n, \tag{4}$$

where *n* is the exponent, $a_n$ is the normalization factor, and $g_e$ is the geometric factor. *n* only depends on the dimension of the materials system, the dispersion relation of the band valley, and the scattering mechanism. $g_e$ is the geometric factor, which is decided by dimension, orientation, anisotropy, etc. To further clarify this, we illustrate the case of a three-dimensional anisotropic band valley, with a dispersion relation of $E(\mathbf{k}) = \hbar^2 k_1^2/2m_1 + \hbar^2 k_2^2/2m_2 + \hbar^2 k_3^2/2m_3$, and a relaxation time function of $\tau_E = \tau_0 (E/k_B T)^s \propto E^s$. The density of states for such a band valley is $D(E) = \sqrt{m_1 m_2 m_3/2}\pi^{-2}\hbar^{-3} E^{0.5} \propto E^{0.5}$, and the transport distribution function for the three principal directions is $\Xi_i(E) = ((m_1 m_2 m_3)^{1/3}/m_i)$ $\cdot (\sqrt{2}\tau_0 (m_1 m_2 m_3)^{1/6}/3\pi^2 \hbar^3 (k_B T)^s) \cdot E^{1+0.5+s}$, where *i*=1, 2 or 3. Thus, we have $n = 1 + 0.5 + s$, $a_n = \sqrt{2}\tau_0 (m_1 m_2 m_3)^{1/6}/3\pi^2 \hbar^3 (k_B T)^s$, and $g_e = (m_1 m_2 m_3)^{1/3}/m_i$ for each direction *i*. Here *s* is only decided by the scattering mechanism, e.g. *s*=0.5 for three-dimensional acoustic phonon scattering. (3) The lattice thermal conductivity can be described by $\kappa_L = g_L \cdot \kappa_L^{bulk}$, where $g_L$ is the lattice geometric factor [11,12].

Similar to $g_e$, $g_L$ is also decided by dimension, orientation, anisotropy, etc., and $g_L = 1$ for an isotropic bulk material.

Under the three above assumptions, we can write $ZT$ as

$$\frac{1}{ZT} = \frac{1}{zt_e} + \frac{1}{zt_L} \frac{1}{(g_e/g_L)a_n} \frac{\kappa_L^{bulk}}{T \cdot k_B^2}, \qquad (5)$$

where $zt_e = (J_0^{[n]} J_2^{[n]} (J_1^{[n]})^{-2} - 1)^{-1}$, $zt_L = (J_1^{[n]})^2 (J_0^{[n]})^{-1}$, $J_r^{[n]} = \int_{x=y}^{\infty} F(x) \cdot (x-y)^n \cdot x^r dx$,

$F(x) = e^x / (1 + e^x)^2$, $y = (E_m - E_f)/k_B T$ and $E_m$ is the energy of the band valley bottom. Thus, we see that $zt_e$ measures the influence of purely electronic behavior to $ZT$, while $zt_L$ scales the influence of the lattice thermal conductivity to $ZT$. $zt_e$ and $zt_L$ are also dimensionless number as $ZT$, which we name as pseudo-$ZT$s. Each pseudo-$ZT$ can be optimized, separately. We also see that the variables that really decides $ZT$ through deciding the pseudo-$ZT$s are: the exponent $n$ and the $zt_L$-pre-factor $(g_e/g_L) \cdot a_n / \kappa_L^{bulk}$, which are independent variables.

Secondly, we examine how dimension of materials, dispersion relation, and carrier scattering mechanism influence $ZT$, by examining how $n$ influence the two pseudo-$ZT$s, $zt_e$ and $zt_L$. We calculated $zt_e$ and $zt_L$ for a single band as a function of Fermi level for different values of $n$, as shown in Fig. 1. We see that (1) $n$ influences $zt_e$ and $zt_L$ in opposite ways. (2) The pre-factor $(g_e/g_L) \cdot a_n / \kappa_L^{bulk}$ decides the portion of the $zt_L$ component. Thus, we have found that as long as an $ZT$-enhancing proposal is associated with choosing or changing the dimension of materials, the dispersion relation, or the carrier scattering mechanism, whether it will enhance $ZT$ or reduce $ZT$ depends on the value of $(g_e/g_L) \cdot a_n / \kappa_L^{bulk}$. For example, the low-dimensionalization proposal only enhances $ZT$ when $(g_e/g_L) \cdot a_n / \kappa_L^{bulk}$ is small so that $ZT$ is dominated by $zt_L$. For large $(g_e/g_L) \cdot a_n / \kappa_L^{bulk}$, where $ZT$ is dominated by $zt_e$, low-dimensionalization actually reduces $ZT$ by reducing $zt_e$.

Thirdly, we examine whether it really enhances *ZT* with proposals related to increasing density of states, by choosing multiband valley materials or by aligning different band edges in energy within the same materials system. We make the examination through an hypothetic procedure, where we can choose to have only one band valley or to have two band valleys contributing to transport, as shown in Fig. 2 (a). *ΔE* denotes the energy difference between band-edges, while *β* denotes the band dissimilarity ratio of transport distribution functions associated with the two band valleys, i.e.

$$\beta = \frac{\Xi^{(2)}(E)}{\Xi^{(1)}(E)} = \frac{g_e^{(2)} \cdot a_n^{(2)}}{g_e^{(1)} \cdot a_n^{(1)}}, \quad (6)$$

where the superscript "$(1)$" and "$(2)$" denote the band valleys. We have calculated the optimization ratio between the one-band-valley cases and the two-band-valley case, as shown in Fig. 2 (b) and (c). The optimization ratio compares the maximum pseudo-*ZT*s after optimized as a function of Fermi level in different cases, which is defined as,

$$\text{Optimization Ratio} = \frac{zt_{max}^{(1)+(2)}}{\max[zt_{max}^{(1)}, zt_{max}^{(2)}]}, \quad (7)$$

where the pseudo-*ZT*s is denoted by a lower case "*zt*", which stands for $zt_e$ or $zt_L$; the superscript "$(1)$", "$(2)$" and "$(1)+(2)$" indicates the band valley (valleys) that is (are) involved in transport; the subscript "$_{max}$" means "maximizing as a function of Fermi level"; $\max[zt_{max}^{(1)}, zt_{max}^{(2)}]$ means the larger one of $zt_{max}^{(1)}$ and $zt_{max}^{(2)}$. Fig. 2 (b) shows that having more bands on the same side of Fermi level does not help $zt_e$: non-zero band-edge displacement *ΔE* can reduce $zt_e$, while band dissimilarity *β* does not influence $zt_e$ significantly. Fig. 2 (c) shows that having more bands on the same side of Fermi level can increases $zt_L$, only if the band-edge displacement *ΔE* is small and the band edge dissimilarity *β* is close to 1. The upper-limit of the optimization ratio for $zt_L$ is the number of band valleys (2 in this two band valley case), i.e. the density of states, which happens only when *ΔE=0* and *β=1*, i.e. the different band valleys are isotropic, and are the same in energy, in dispersion form and in scattering mechanism. Thus, Fig. 2 (b) and (c) ex-plain why proposals associated with increasing density of states at the band edges can sometimes enhance *ZT*, and also explain how density of states should be increased to enhance $zt_L$ without jeopardizing $zt_e$, i.e. having the different band valleys as similar to each other as possible both in energy and in transportation distribution function.

Fourthly, we will show that the asymmetry of transportation functions between the conduction band and the valence band is a variable that enhances both $zt_e$ and $zt_L$ at the same time. Without loss of generality, we consider an hypothetic situation where we have one conduction band valley and one valence band valley that are separated by the band gap $E_g$, as shown in Fig. 3 (a). The asymmetry degree $\gamma$ is defined as

$$\gamma = \frac{\Xi^{(v)}(E)}{\Xi^{(c)}(E)} = \frac{g_e^{(v)} \cdot a_n^{(v)}}{g_e^{(c)} \cdot a_n^{(c)}}, \tag{8}$$

which is similar to the definition of $\beta$, except "(c)" ("(v)") here denotes the conduction (valence) band. In Fig. 3 (b) and (c), we have illustrated how $zt_e$ and $zt_L$ monotonically increases as a function $\gamma$ for the case where $E_g=20k_BT$, respectively. Other values of band gap exhibit the same tendency that pseudo-*ZT*s monotonically increases as a function of $\gamma$. From Fig. 3 (b) and (c), we see that enhancing *ZT* prefers holes to be as different from electrons as possible, i.e. *ZT* benefits from large asymmetry degree $\gamma$. This explains why materials with both heavy holes and light electrons, such as topological insulator, quasi-Dirac-cones, or narrow-band-gaps, are always good thermoelectric materials, i.e. these materials can provide large asymmetry degree $\gamma$. This also explains why some nano-composites systems have been reported to have high *ZT*, because the holes-screening grain-boundaries in nano-composites materials are effectively making holes much heavier than electrons, which results in a large asymmetry degree of $\gamma$.

Lastly, we have found that large band gap can enhance both pseudo-*ZT*s, until the band gap becomes highly correlated with lattice thermal conductivity. We still use the hypothetic system in Fig. 3 (a), except that we now fix $\gamma$ and examine $zt_e$ and $zt_L$ as a function of $E_g$. Without loss of generality, we illus-trate the case where $\gamma=1$, as shown in Fig. 4 (a) and (b). Other values of $\gamma$ exhibit the similar tendency that both pseudo-*ZT*s increase as a function of $E_g$. Fig. 4 has dissolved the traditional "fare" of large band gap, which implies small carrier concentration. Our calculations in Fig. 4 clearly shows that both of the pseudo-*ZT*s increases as a function of $E_g$. However, it is worth to keep alert that the Fermi level of large band gap insulators could become difficult to tune through doping. Furthermore, large band gap may be positively correlated with $\kappa_L^{bulk}$, i.e. large band gap insulators usually have long inter-atomic distances and results in large lattice thermal conductivity, which compromises the enhanced $zt_L$ value by reduced $(g_e/g_L) \cdot a_n / \kappa_L^{bulk}$ in Eq. (5).

In conclusion, we have pointed out that the widely used expression of *ZT* as a function of $S$, $\sigma$, and $\kappa$ is the obstacle of solving the *ZT* enhancement problem, and provided a more efficient approach to solve it. We have proposed the idea of splitting *ZT* into two pseudo-*ZT*s, $zt_e$ and $zt_L$, such that the influence on *ZT* from purely electronic behavior and from lattice thermal conductivity can be separately treated. Based on this concept of pseudo-*ZT*s, we have found that instead of $S$, $\sigma$, or $\kappa$, the variables that we can use to solve the mathematical problem of *ZT* optimization are: $n$ and $(g_e/g_L) \cdot a_n / \kappa_L^{bulk}$ for single-band properties, $\beta$ and $\Delta E$ for the multiband-relation, and $\gamma$ and $E_g$ for the electron-hole relation.

We have pointed out that all the enhancing *ZT* proposals that are related to changing or choosing the dimension of materials, the forms of dispersion relation, or the carrier scattering mechanism will change $zt_e$ and $zt_L$ in opposite ways. Therefore, these kinds of enhancing *ZT* proposals must be carried out with caution. We have illustrated why the proposal of low-dimensionalization does not always enhance *ZT*: low-dimensionalization increases $zt_e$ while decreases $zt_L$, so it only works when $(g_e/g_L) \cdot a_n / \kappa_L^{bulk}$ is small and *ZT* is dominated by $zt_L$. For large $(g_e/g_L) \cdot a_n / \kappa_L^{bulk}$ cases, where *ZT* is dominated by $zt_e$, low-dimensionalization actually reduces *ZT*. We have also pointed out that pursuing large density of states by choosing multi-band-valley materials is not always enhancing *ZT*. *ZT* is only benefited when the different band valleys on the same side of Fermi level are similar to each other. Most importantly, we have found out that the asymmetry degree between electrons and holes, and the band gaps are the parameters that really enhances *ZT*, by increasing both of the pseudo-*ZT*s at the same time.

Finally, we have proposed the standard principles of *ZT* enhancing: (1) Reducing (Increasing) *n* by changing or choosing the dimension of materials, the forms of dispersion relation, or the carrier scattering mechanism for small (large) $(g_e/g_L) \cdot a_n / \kappa_L^{bulk}$. (2) Increasing density of states on the same side of Fermi level while keeping different band valleys as similar to each other as possible. (3) Keeping the conduction band and the valence band as asymmetry as possible. (4) Having large band gap as long as it does not significantly influence the doping capability and the lattice thermal conductivity.

**Figures:**

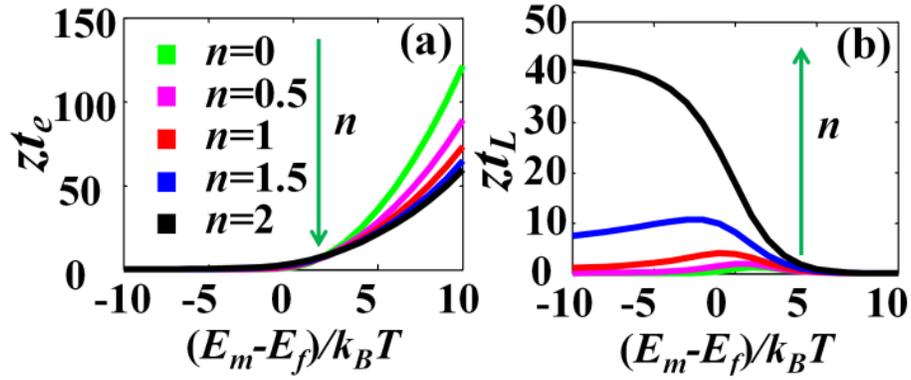

**Figure 1**: (a) $zt_e$ and (b) $zt_L$ for a single band as a function of Fermi level for different values of $n$, where $E_m$ is the bottom of band and $E_f$ is the Fermi level. It clearly shows that $zt_e$ increases, while $zt_e$ decreases with n, which is the variable that reflects the dimension of materials, dispersion relation form, and scattering mechanism. Therefore, we see that all the *ZT* enhancing proposals by changing or choosing dimension, dispersion relation form or scattering mechanism, actually affect the two pseudo-*ZT*s in opposite directions.

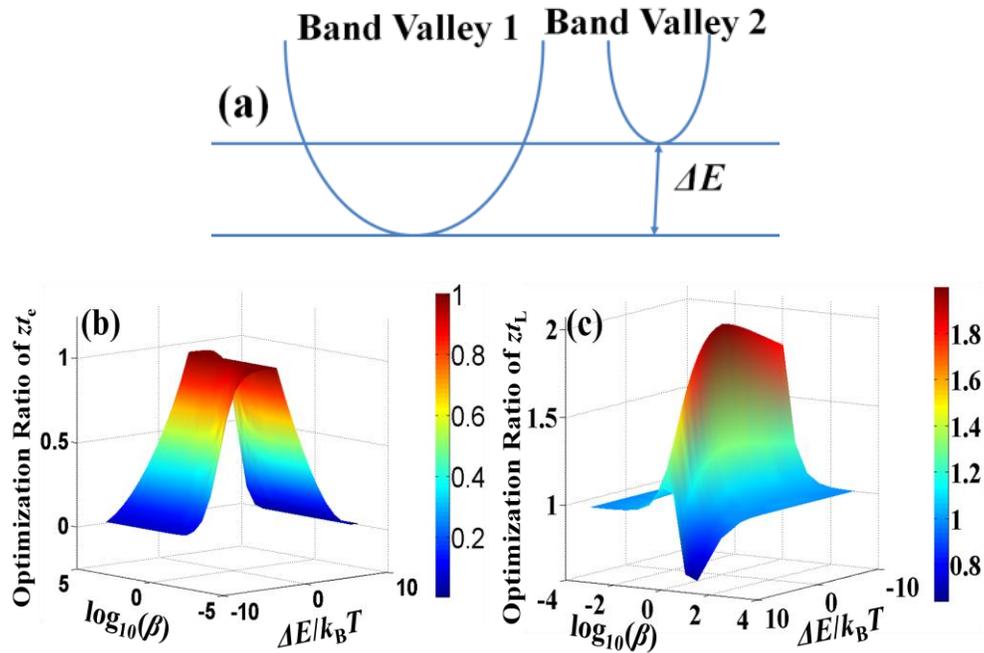

**Figure 2**: The effect of having more bands on pseudo-*ZT*s. (a) Scheme of Band Valley 1 and Band Valley 2. The optimization ratio of (b) $zt_e$ and (c) $zt_L$ between the hypothetic case of having two band valleys and the hypothetic cases of having one band valley contributing to thermoelectric transport are shown as a function of band-edge displacement $\Delta E$ and band dissimilarity $\beta$. Different values of *n* give the same tendency of optimization ratio changes as a function of $\Delta E$ and $\beta$, and $n=2$ is illustrated in this figure.

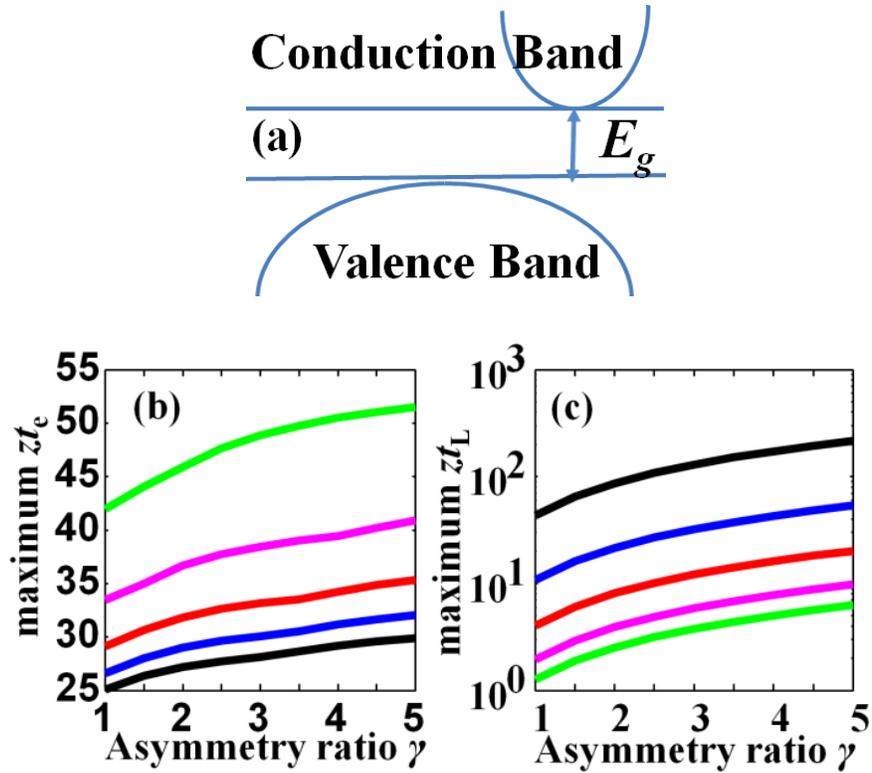

**Figure 3**: How band asymmetry $\gamma$ enhances both pseudo-*ZT*s. (a) Scheme of the conduction band edge and the valence band edge. Different $n$ values are specified by colors that are identically defined in the legend of Fig. 1. Both (b) $zt_e$ and (c) $zt_L$ increases as a function of the band asymmetry degree $\gamma$. $\gamma$ is a variable that really enhances *ZT* by increasing both pseudo-*ZT*s at the same time.

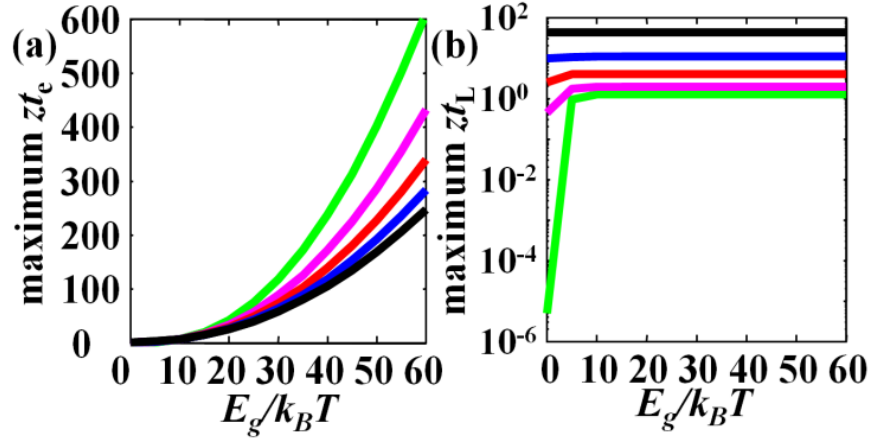

**Figure 4**: How band gap benefits both pseudo-*ZT*s. Different *n* values are specified by colors that are identically defined in the legend of Fig. 1. Both (a) $zt_e$ and (b) $zt_L$ increases as a function of the band gap $E_g$. $E_g$ is another variable that really enhances *ZT* by increasing both pseudo-*ZT*s at the same time. (b) shows that increasing the band gap helps increasing $zt_L$ very obviously when $E_g$ is smaller than $10k_BT$, and also does not decreases $zt_L$ when $E_g$ is larger than $10k_BT$. However, it is worth to keep alert that when the Fermi level of large band gap insulators becomes difficult to change through doping, or when large band gap become positively correlated with $\kappa_L^{bulk}$, i.e. when large band gap insulators have large inter-atomic distances that results in large lattice thermal conductivity, the enhancement of $zt_L$ might be compromised by the reduction of $(g_e/g_L) \cdot a_n / \kappa_L^{bulk}$ in Eq. (5).

**References:**

[1] T. M. Tritt, Science **283**, 804 (1999).

[2] L. Hicks, and M. Dresselhaus, Phys. Rev. B **47**, 12727 (1993).

[3] G. Mahan, and J. Sofo, Proceedings of the National Academy of Sciences **93**, 7436 (1996).

[4] Y. Lin, and M. Dresselhaus, Phys. Rev. B **68**, 75304 (2003).

[5] J. P. Heremans *et al.*, Science **321**, 554 (2008).

[6] B. Poudel *et al.*, Science **320**, 634 (2008).

[7] D. Parker, X. Chen, and D. J. Singh, Phys. Rev. Lett. **110**, 146601 (2013).

[8] H. J. Goldsmid, *Introduction to thermoelectricity* (Springer, 2009), Vol. 121.

[9] N. W. Ashcroft, and N. D. Mermin, *Solid state physics* (Saunders College, 1976).

**Acknowledgements:**

We acknowledge the support from AFOSR MURI Grant Number FA9550-10-1-0533, Subaward 60028687.

**Author Contributions:**

S. T. conceived the idea, made the calculations and wrote the paper. M. S. D. provided the research platform and funding, and helped in writing the paper.

**Competing financial interests:**

The authors declare no competing financial interests.

1   Tritt, T. M. Holey and unholey semiconductors. *Science* **283**, 804-805 (1999).

2   Hicks, L. & Dresselhaus, M. Effect of quantum-well structures on the thermoelectric figure of merit. *Phys. Rev. B* **47**, 12727-12731 (1993).

3   Mahan, G. & Sofo, J. The best thermoelectric. *Proceedings of the National Academy of Sciences* **93**, 7436-7439 (1996).

4   Venkatasubramanian, R., Siivola, E., Colpitts, T. & O'quinn, B. Thin-film thermoelectric devices with high room-temperature figures of merit. *Nature (London)* **413**, 597-602 (2001).

5   Harman, T., Taylor, P., Walsh, M. & LaForge, B. Quantum dot superlattice thermoelectric materials and devices. *Science* **297**, 2229-2232 (2002).

6   Lin, Y. & Dresselhaus, M. Thermoelectric properties of superlattice nanowires. *Phys. Rev. B* **68**, 75304 (2003).

7   Heremans, J. P. *et al.* Enhancement of thermoelectric efficiency in PbTe by distortion of the electronic density of states. *Science* **321**, 554-557 (2008).

8   Poudel, B. *et al.* High-thermoelectric performance of nanostructured bismuth antimony telluride bulk alloys. *Science* **320**, 634-638 (2008).

9   Biswas, K. *et al.* High-performance bulk thermoelectrics with all-scale hierarchical architectures. *Nature (London)* **489**, 414-418 (2012).

10  Parker, D., Chen, X. & Singh, D. J. High Three-Dimensional Thermoelectric Performance from Low-Dimensional Bands. *Phys. Rev. Lett.* **110**, 146601 (2013).

11  Goldsmid, H. J. *Introduction to thermoelectricity*. Vol. 121 (Springer, 2009).

12  Ashcroft, N. W. & Mermin, N. D. *Solid state physics*.  (Saunders College, 1976).